\documentclass[a4paper,11pt]{article}
\usepackage{pos}
\usepackage{float}
\usepackage[symbol]{footmisc}
\usepackage{setspace}
\renewcommand{\thefootnote}{\fnsymbol{footnote}}
\newcommand{\KK}{${\cal KK}$}
\def\rQCED{{\rm QCED}}
\newcommand{\qb}{{\bar{q}}}
\newcommand{\sfac}{\mathfrak{s}}
\title{On Amplitude-Based IR-Improvement in Precision LHC$/$FCC Physics}
\centerline{BU-HEPP-26-02, 11/2026}
\author*[a,b]{B.F.L. Ward}
\author[c]{S. Jadach\footnote[2]{Deceased.}} 
\author[d]{W. Placzek}
\author[a]{S. Roy}
\author[d]{A. Siodmok}
\author[c]{M. Skrzypek}
\author[c]{Z. A. Was}
\author[e]{S.A. Yost}

\affiliation[a]{Department of Physics and Astronomy, Baylor University,\\
  Waco, TX, USA}

\affiliation[b]{RISC, Johannes Kepler University,\\
Linz, Austria}

\affiliation[c]{Institute of Nuclear Physics,\\
Krakow, PL}

\affiliation[d]{Institute of Applied Computer Science, Jagiellonian University,\\
Krakow, PL}

\affiliation[e]{Department of Physics, The Citadel,\\
  Charleston, SC, USA}

\emailAdd{bfl\_ward@baylor.edu}
\emailAdd{wieslaw.placzek@uj.edu.pl}
\emailAdd{suvadeep\_roy1@baylor.edu}
\emailAdd{a.siodmok@cern.ch}
\emailAdd{maciej.skrzypek@ifj.edu.pl}
\emailAdd{zbigniew.was@ifj.edu.pl}
\emailAdd{yosts1@citadel.edu}

\abstract{Using IR-improvement of unintegrable singularities in the infrared regime via amplitude-based resummation in $QED \times QCD \subset SU(2)_L \times U_1 \times SU(3)_c$ we present new results in the context of precision LHC$/$FCC physics. We focus on specific examples, such as the interplay between parton showers and the removal of QED contamination in PDF’s evolved from data at $Q_0^2\sim 2 GeV^2$  and used in evaluating precision observables in $pp\rightarrow Z + X \rightarrow \ell\bar\ell +  X’$. We discuss attendant new issues.}

\FullConference{43rd International Conference on High Energy Physics (ICHEP2026)\\
30 July to 5 August , 2026\\
Natal, Brazil\\}

\begin{document}
\maketitle

\baselineskip=10pt
\section{Introduction}
What is relatively safe to say is that the long-term expectations for precision theory are determined by future accelerators. Here, we list the following: HL-LHC, FCC, CLIC, ILC, CEPC, SPPC, SSC-RESTART?  If we focus on FCC for illustration, factors of improvement 
from $\sim 5$  to $\sim 100$ are needed from theory. As we see in Ref.~\cite{fabiola-1-10-23}, at world-leading laboratories such as CERN the need for precision theory for the success of future collider physics programs is recognized. The presentation in the latter reference shows possible future options for CERN featuring the FCC with the important role of higher-order calculations, with their attendant computer algebraic methodology, for its background processes as a theory highlight -- presumably, the funding agencies appreciate the implied connection. 
We can see from Ref.~\cite{grojean-2-13-24} the factor of $\sim 5\;{\rm to}\; 10$ advantage in reach of the FCCee compared to
the ${\rm ILC}_{250}$ in precision EW tests of the Standard Theory of elementary particles -- there will be $\sim$ 50 times better precision relative to LEP/SLD. Resummation is key to such calculations/analyses in many cases. In what follows, we discuss amplitude-based resummation following
the YFS~\cite{yfs:1961}  MC methodology~\cite{krkepiph-2024,mttd2025-bflw}.\par
For reference, we note that YFS methods are exact in the infrared limit but treat the collinear logs via partial resummation with a remainder
perturbatively included in the $\bar{\beta}_n$ residuals. In contrast, DGLAP-based collinear factorization treats the collinear logs to all
orders but has a non-exact IR limit -- see Ref.~\cite{bertone-2022} and references therein. In what follows, we present new results for precision collider physics
based on the usual YFS methods with an eye toward their new collinearly improved counterparts~\cite{Jadach:2023cl1}. A key point to keep in mind is that exact amplitude-based resummation realized on an event-by-event basis implies enhanced precision for a given level of exactness:
LO, NLO, NNLO, NNNLO, ... . This is essential for future precision physics as exemplified by CERN, wherein we stress that computer algebraic methods are essential.
\par
Our discussion proceeds as follows. In the next section we give a brief of review of YFS exact amplitude-based resummation theory, as it is not generally familiar. In Section 3, we
turn to new perspectives for precision collider physics with an eye toward LHC, FCC, CEPC, SPPC, LCF, ILC, and CLIC. Section 4 contains our concluding discussion.\par
\section{Review of YFS Exact Amplitude-Based Resummation}
It is still true that our YFS~\cite{yfs:1961} exact amplitude-based resummation theory, especially its CEEX~\cite{Jadach:2000ir} realization, is generally not familiar in our field. We present in this section a brief recapitulation of the theory accordingly before turning to some new results based on it in the next section.\par
The theory is exhibited by the following master formula:
{\small
\begin{eqnarray}
&d\bar\sigma_{\rm res} = e^{\rm SUM_{IR}(QCED)}
   \sum_{{n,m}=0}^\infty\frac{1}{n!m!}\int\prod_{j_1=1}^n\frac{d^3k_{j_1}}{k_{j_1}} \cr
&\prod_{j_2=1}^m\frac{d^3{k'}_{j_2}}{{k'}_{j_2}}
\int\frac{d^4y}{(2\pi)^4}e^{iy\cdot(p_1+q_1-p_2-q_2-\sum k_{j_1}-\sum {k'}_{j_2})+
D_\rQCED} \cr
&{\tilde{\bar\beta}_{n,m}(k_1,\ldots,k_n;k'_1,\ldots,k'_m)}\frac{d^3p_2}{p_2^{\,0}}\frac{d^3q_2}{q_2^{\,0}},
\label{subp15b}
\end{eqnarray}}
where the {\em new} (YFS-style) residuals   
{$\tilde{\bar\beta}_{n,m}(k_1,\ldots,k_n;k'_1,\ldots,k'_m)$}, which have {$n$} hard gluons and {$m$} hard photons, contain the QCD color correlations and the remainder of the QCD exponent from Ref.~\cite{Gatheral:1983}. The new residuals and the  infrared functions ${\rm SUM_{IR}(QCED)}$ and ${ D_\rQCED}$ are defined in Ref.~\cite{mcnlo-hwiri,mcnlo-hwiri1}.  For reference, we point out that, to meet the precision tags required for the FCCee, we would need the $\tilde{\bar\beta}_{n,m}$ exact in ${\cal O}\left(\frac{\alpha}{\pi}, \frac{\alpha}{\pi}L, \left(\frac{\alpha}{\pi}\right)^2, \left(\frac{\alpha}{\pi}\right)^2L, \left(\frac{\alpha}{\pi}\right)^2L^2, \left(\frac{\alpha}{\pi}\right)^3L^2, \left(\frac{\alpha}{\pi}\right)^3L^3,
 \left(\frac{\alpha}{\pi}\right)^4L^4\right)$ where $L = \ln\frac{s}{m_e^2}$ is the usual big log in an obvious notation.
This further implies that the role of computer algebraic methods~\cite{blum-schneider-2023,kalmykov:2020} in evaluating the respective Feynman diagrams will be significant. As explained in Ref.~\cite{mcnlo-hwiri,mcnlo-hwiri1}, parton shower/ME matching induces the replacements {$\tilde{\bar\beta}_{n,m}\rightarrow \hat{\tilde{\bar\beta}}_{n,m}$}, which allow us to connect with  MC@NLO~\cite{mcnlo,mcnlo1}, via the basic formula{\small
\begin{equation}
{d\sigma} =\sum_{i,j}\int dx_1dx_2{F_i(x_1)F_j(x_2)} d\hat\sigma_{\rm res}(x_1x_2s).
\label{bscfrla}
\end{equation}}
\par
We have used eq.(\ref{subp15b}) to obtain new results in precision LHC and FCC physics. One of us (BFLW), by extending (See Ref.~\cite{ijmpa2018} and references therein.)  eq.(\ref{subp15b}) to general relativity, has developed a new approach to quantum gravity.  As we illustrate in the next Section, in each of our applications new results are accompanied with new perspectives.\par

\section{New Perspectives for Precision Collider Physics: LHC, FCC, CEPC, SPPC, LCF, ILC, CLIC}
\begin{figure}[h!]
\begin{center}
\setlength{\unitlength}{1in}
\begin{picture}(5.5,2.0)(0,0)
\put(0.5,0){\includegraphics[width=4.5in]{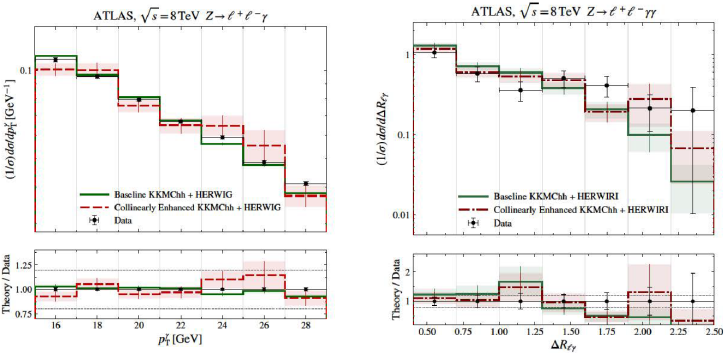}}
\end{picture}
\end{center}
\vspace{-5mm}
\caption{Comparison of the ATLAS data on $Z/\gamma$ production at $8$ TeV with the collinearly enhanced (red dashed curves) and non-enhanced (green solid curves) versions of KKMChh for the $p^\gamma_T$ and $R_{\ell\gamma}$ observables as defined in Ref.~\cite{atlas-epjc-zg:2024}.}
\label{fig3}
\end{figure}

Due to the  realization of eq.(\ref{subp15b}) with the collinear enhancement in Ref.~\cite{Jadach:2023cl1} in the MC event generator \KK{MC}-hh~\cite{kkmchh1-sh} in v. 1.312 by five of us (SJ, BFLW, ZAW, SAY,SR), we have a new perspective on the expectations for precision physics for the Standard Theory
EW interactions at HL-LHC. This is illustrated here with the plots in Fig.~\ref{fig3} in the ATLAS analysis~\cite{atlas-epjc-zg:2024} of $Z\gamma$ production at 8 TeV.
Higher statistics comparisons expected at
HL-LHC would presumably allow a definitive check on the collinear enhancement improvement in the respective precision. The latter also would be important for
the physics program of the FCC.\par
\begin{figure}[h!]
\begin{center}
\setlength{\unitlength}{1in}
\begin{picture}(6,2.0)(0,0)
\put(0.6,-.15){\includegraphics[width=4.7in]{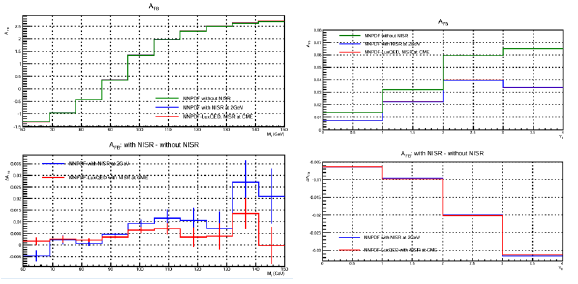}}
\end{picture}
\end{center}
\vspace{-5mm}
\caption{The first plot on the left shows the dependence of $A_{FB}$ on the final lepton (muon) invariant mass $M_{\ell\ell}$ for the three cases under study: NNPDF-3.1 without NISR (green), NNPDF-3.1 with NISR at 2 GeV (blue), and NNPDF-3.1-LuxQED (red) with NISR at the generated CM energy. The second plot on the left shows the differences between each version with NISR and the version without  NISR. The shifts from adding NISR are mostly comparable for both versions. The plots on the right give the same information for $Y_{\ell\ell}$.}
\label{fig4}
\end{figure}

Toward the important issue of the effect of QED contamination in non-QED PDFs ~\cite{jad-yost-afb,ichep2022-say,sjetaltoappear}, in a new perspective we use Negative ISR (NISR) evolution to address directly the size of this contamination. We have the representation of the cross section, using a standard notation for PDFs and cross sections, 
\begin{equation}
\begin{split}
\sigma(s)&=
\frac{3}{4}\pi\sigma_0(s)\!\!\!
\sum_{q=u,d,s,c,b} \int d\hat{x}\;  dz dr dt \; \int dx_q dx_\qb\; 
\delta(\hat{x}-x_qx_\qb z t)
\\&\times
f^{h_1}_q(   s\hat{x}, x_q) 
f^{h_2}_\qb( s\hat{x}, x_\qb) \;
 \rho_I^{(0)}\big(\gamma_{Iq}(s\hat{x}/m_q^2),z\big)\; 
 \rho_I^{(2)}\big(-\gamma_{Iq}(Q_0^2/m_q^2),t\big)\; 
\\&\times
\sigma^{Born}_{q\qb}(s\hat{x}z)\;
\langle W_{MC} \rangle,
\label{eq:kkhhsigmaPru}
\end{split}
\end{equation}
which includes an extra convolution with the well known second order exponentiated ISR
``radiator function'' $\rho_I^{(2)} $ with the negative evolution time argument
$-\gamma_{Iq}(Q_0^2/m_q^2)$ defined in Ref.~\cite{jad-yost-afb}. This removes the  QED below $Q_0$. This is exhibited in Fig.~\ref{fig4} from Ref.~\cite{ichep2022-say}  
for the $A_{FB}$ and $Y_{\ell\ell}$. As explained in the caption, NISR shows the effect of QED contamination in non-QED PDFs is below the errors on the PDFs themselves, in agreement with the arguments in Ref.~\cite{kkmchh2}.\par
\begin{figure}[h!]
\begin{center}
\setlength{\unitlength}{1in}
\begin{picture}(6.0,2.55)(0,0)
\put(0.7,-,1){\includegraphics[width=4.5in]{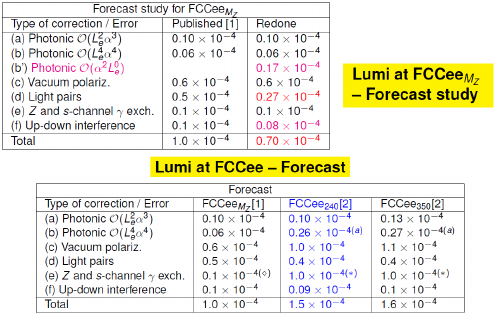}}
\end{picture}
\end{center}
\vspace{-5mm}
\caption{Current purview on luminosity theory errors: top, FCC-ee at $M_Z$; bottom, proposed higher energy FCCee collider runs.}
\label{fig5}
\end{figure}

In view of the planned EW/Higgs factories, five of us (SJ, WP, MS, BFLW, SAY) have discussed in Refs.~\cite{Jadach:2018jjo,Jadach:2021ayv-sh,fcc2023wkshpms,skrzypek-2023mttd,Placzek:2026hvx} the new perspectives for the BHLUMI~\cite{bhlumi4:1996} luminosity theory error. In Fig.~\ref{fig5}~\cite{fcc2023wkshpms}, to illustrate this new perspective,  we show the current purview for the FCC-ee at $M_Z$ and for the proposed higher energy FCCee collider runs. 
We note the improvements at $M_Z$ shown in Fig.~\ref{fig5}(top) to 0.007\% and that the use of the results in Ref.~\cite{fjeger-fccwksp2019} together with lattice methods~\cite{latt1,latt2,latt3}
opens the possibility that item (c) in Fig~\ref{fig5}(top) could be reduced by a factor of $\sim 6$\footnote{The formula to be studied is $\Delta\alpha_{had}(t)=\Delta\alpha_{had}(-Q^2_0)|_{lat}+[\Delta\alpha_{had}(t)-\Delta\alpha_{had}(-Q^2_0)]|_{pQCDAdler}$ with {\it lat} denoting the methods of Refs.~\cite{latt1,latt2,latt3} and {\it pQCDAdler} denoting the methods of Ref.~\cite{fjeger-fccwksp2019}.}.  See also Ref.~\cite{Jegerlehner:2026zwk}. In addition, see  Ref.~\cite{bflw-ichep2026} for discussion of the possibility to reach $10^{-6}$ lumi precision.\par
One of the further applications of eq.(\ref{subp15b}) is in the realization and improvement of KKMCee~\cite{Jadach:2022vf} to the C++ version 5.00.2. KKMC~\cite{Jadach:1999vf-sh} has split into two branches, one is KKMCee and the other is KKMChh~\cite{Jadach:2016kh,kkmchh1-sh} with the same EW physics as KKMCee but for quark-anti-quark collisions. The C++ paradigm in KKMCee therefore carries over to one for KKMChh. This realization is in progress with a dominant role already played by the late Prof. S. Jadach: benchmarks have been passed, etc. For example, just as we have a 5D KKeeFoam for semi-analytical cross-check of KKMCee, so too do we have 5D KKhhFoam for the similar cross-check of the KKMChh C++ version. The results are still in development, but stay tuned!\par
Given the higher precision requirements for HL-LHC/FCC physics, we are motivated to improve the collinear limit of YFS theory. In particular, it is known~\cite{Jadach:2023cl1} that, for the process $e^+(p_2)\;e^-(p_1)\rightarrow \bar{f}(p_4)\;f(p_3)$, in the usual YFS theory the s-channel virtual infrared function $B$ resums (exponentiates) the non-infrared term
$\frac{1}{2}Q_e^2{\alpha\over\pi} L$ whereas the methods in Ref.~\cite{gribv-lptv:1972} show that the term $\frac{3}{2}Q_e^2{\alpha\over\pi} L$ exponentiates -- see also Refs.~\cite{frixione-2019,bertone-2019,frixione-2021,bertone-2022} for recent developments in the attendant collinear factorization approach. Three of us (SJ, BFLW, ZAW) in Ref.~\cite{Jadach:2023cl1} have shown that the suggested collinear enhancement of YFS theory which exponentiates the term $\frac{3}{2}Q_e^2{\alpha\over\pi} L$ does exist.\par
More precisely, we show that the virtual infrared function $B$ in the s-channel and the corresponding real infrared function $\tilde{B}$ can be extended to {\small
\begin{equation}
\begin{split}
B_{CL}&\equiv B+{\bf \Delta B}\\
          &= \int {d^4k\over k^2} {i\over (2\pi)^3} 
                       \bigg[\bigg( {2p-k \over 2kp-k^2} - {2q+k \over 2kq+k^2} \bigg)^2{\bf -\frac{4pk-4qk}{(2pk-k^2)(2qk+k^2)}}\bigg],
\end{split}
\end{equation}}
 {\small
\begin{equation}
\begin{split}
\tilde{B}_{CL} &\equiv \tilde{B}+{\bf \Delta\tilde{B}}\\
 &=\frac{-1}{8\pi^2}\int\frac{d^3k}{k_0}\bigg\{\large(\frac{p_1}{kp_1} - \frac{p_2}{kp_2}\large)^2 +{\bf \frac{1}{kp_1}\large(2 -\frac{kp_2}{p_1p_2}\large)}
                                          +{\bf \frac{1}{kp_2}\large(2 -\frac{kp_1}{p_1p_2}\large)}\bigg\},
\end{split}
\label{eq-real2}
\end{equation}}
\noindent where the extensions are indicated in boldface in an obvious notation. These extensions leave the YFS infrared algebra unaffected while the $B_{CL}$ does exponentiate the entire $\frac{3}{2}Q_e^2{\alpha\over\pi} L$ term and the $\tilde{B}_{CL}$ does carry the respective collinear big log of the exact result in Ref.~\cite{berends-neerver-burgers:1988} in the soft regime.\par 
In the CEEX case, the corresponding collinear extension of the soft eikonal amplitude factor defined in Ref.~\cite{ceex2:1999sh}
for the photon polarization $\sigma$ and $e^-$  helicity $\sigma'$ is given by {\small
\begin{equation}
\begin{split}
\sfac_{CL,\sigma}(k) = \sqrt{2}Q_ee\bigg[-\sqrt{\frac{p_1\zeta}{k\zeta}}\frac{<k\sigma|\hat{p}_1 -\sigma>}{2p_1k}
    +{\bf \delta_{\sigma'\;-\sigma}\sqrt{\frac{k\zeta}{p_1\zeta}}\frac{<k\sigma|\hat{p}_1 \sigma'>}{2p_1k}}\\
    + \sqrt{\frac{p_2\zeta}{k\zeta}}\frac{<k\sigma|\hat{p}_2 -\sigma>}{2p_2k}+{\bf \delta_{\sigma' \sigma}\sqrt{\frac{k\zeta}{p_2\zeta}}\frac{<\hat{p}_2 \sigma'|k -\sigma>}{2p_2k}}\bigg],
\end{split}
\label{eq-real4}
\end{equation}}
where from Ref.~\cite{ceex2:1999sh} $\zeta\equiv (1,1,0,0)$ for our choice for the respective auxiliary vector in our Global Positioning of Spin (GPS)~\cite{gps:1998} spinor conventions with the consequent definition 
$\hat{p}= p - \zeta m^2/(2\zeta p)$
for any four vector $p$ with $p^2 = m^2.$ The collinear extension terms are again indicated in boldface.
Upon taking the modulus squared of $\sfac_{CL,\sigma}(k)$ we see that the extra
non-IR divergent contributions reproduce the known collinear big
log contribution which is missed by the usual YFS algebra. KKMChh 1.312 now carries the attendant collinearly enhanced algebra.
\par
These extended infrared functions are expected to give in general a higher precision for a given level of exactness~\cite{bflwetaltoappear}.\par
\section{Summary}
We conclude that amplitude-based resummation allows improved control of the infrared (IR) and
collinear (CL) limits. In general, MC realizations are needed for current and future precision collider
physics. The new, collinearly enhanced soft functions imply a higher level of accuracy for
a given level of exactness in the IR-finite YFS hard photon residuals. This yields an enhanced toolbox available to extend the (CEEX) YFS MC method to the other important processes at present and future colliders. We emphasize that some New Physics may hang in the balance at the LHC, FCC, and other future colliders!
\section*{Acknowledgments}
Two of us (BFLW, ZAW) would like to thank Prof. U. Wiedemann for the support and kind hospitality of the CERN TH Department. This work has been supported by the National Science Centre (NCN), Poland under the grant No. 2023/50/A/ST2/00224. S.A.Yost was supported in part by a grant from the Citadel Foundation.\par
\baselineskip=10pt
\bibliography{Tauola_interface_design}{}

@article{Jegerlehner:2026zwk,
    author = "Jegerlehner, Fred",
    title = "{Lepton Magnetic Moments: What They Tell Us}",
    eprint = "2601.08101",
    archivePrefix = "arXiv",
    primaryClass = "hep-ph",
    reportNumber = "HU-EP-26/03",
    doi = "10.5506/APhysPolBSupp.19.2-A1",
    journal = "Acta Phys. Polon. Supp.",
    volume = "19",
    number = "2",
    pages = "2--A1",
    year = "2026"
}

@article{Placzek:2026hvx,
    author = "P{\l}aczek, Wies{\l}aw and Skrzypek, Maciej and Ward, Bennie F. L. and Yost, Scott A.",
    title = "{Bhabha Scattering at Future Colliders with BHLUMI/BHWIDE}",
    eprint = "2601.15265",
    archivePrefix = "arXiv",
    primaryClass = "hep-ph",
    doi = "10.5506/APhysPolBSupp.19.2-A2",
    journal = "Acta Phys. Polon. Supp.",
    volume = "19",
    number = "2",
    pages = "2--A2",
    year = "2026"
}

@inproceedings{skrzypek-2023mttd,
    author = "Skrzypek, M. and  P\l{}aczek, W. and Ward, B. F.L.  and Yost, S.A.",
    title = "{How well could we calculate luminosity at FCCee?}",
    booktitle = "{Proc. XLV MTTD Conference, Ustro\'n, PL, 2023}",
    eprint = "",
    archivePrefix = "",
    primaryClass = "",
    reportNumber = "BU-HEPP-23-03",
    doi = "",
    publisher = "Acta Phys. Pol. B Proc. Suppl.",
    address = "",
    series = "",
    volume = "17",
    pages = "2-A4",
    month = "",
    year = "2024"
}

@article{mttd2025-bflw,
      author         = "Ward, B.F.L.",
      title          = "{YFS MC Approach to Precision Theory for Collider Physics: Origin, Development, and Outlook}",
      journal        = "Symmetry",
      volume         = "18",
      pages          = "1286",
      doi            = "10.3390/sym18081286",
      year           = "2026",
      eprint         = "",
      archivePrefix  = "",
      primaryClass   = "",
      reportNumber   = "",
      SLACcitation   = "",
}

@inproceedings{krkepiph-2024,
    author = "Ward, B. F.L.",
    title = "{YFS Exponentiation – Gate to Precision Accelerator Physics Experiments}",
    booktitle = "{Proc. 2024 Krakow Epiphany Conference, Krakow, PL, 2024}",
    eprint = "",
    archivePrefix = "",
    primaryClass = "",
    reportNumber = "BU-HEPP-24-01",
    doi = "",
    publisher = "Acta Phys. Pol. B Proc. Suppl.",
    address = "",
    series = "",
    volume = "17",
    pages = "5-A2",
    month = "",
    year = "2024"
}

@article{Jadach:2021ayv-sh,
    author = "Jadach, S. and P\l{}aczek, W. and Skrzypek, M. and Ward, B. F. L.",
    title = "{Study of theoretical luminosity precision for electron colliders at higher energies}",
    doi = "10.1140/epjc/s10052-021-09860-9",
    journal = "Eur. Phys. J. C",
    volume = "81",
    pages = "1047",
    year = "2021"
}

@article{blum-schneider-2023,
    author = "Bluemlein, J and Fadeev, N. and Schneider, C.",
    title = "",
    doi = "",
    journal = "Commun. Comput. Algebra",
    volume = "57",
    pages = "224",
    year = "2023"
}

@inproceedings{kalmykov:2020,
    author = "Kalmykov, M. and others",
    title = "{Hypergeometric Functions and Feynman Diagrams}",
    booktitle = "{Proc. of Workshop on Anti-Differentiation and the Calculation of Feynman Amplitudes}",
    eprint = "2012.14492",
    archivePrefix = "arXiv",
    primaryClass = "hep-ph",
    reportNumber = "BU-HEPP-20-09",
    doi = "",
    publisher = "Springer",
    address = "Berlin",
    series = "",
    volume = "",
    pages    = "189",
    month = "",
    year = "2021"
}

@article{Gatheral:1983,
    author = "Gatheral, J.G.M.",
    title = "{Exponentiation of eikonal cross sections in nonAbelian gauge theories}",
    doi = "10.1016/0370-2693(83)90112-0",
    journal = "Phys. Lett.",
    volume = "133B",
    pages = "90",
    year = "1983"
}

@article{Jadach:2018jjo,
    author = "Jadach, S. and P\l{}aczek, W. and Skrzypek, M. and Ward, B. F. L. and Yost, S. A.",
    title = "{The path to 0.01\% theoretical luminosity precision for the FCC-ee}",
    eprint = "1812.01004",
    archivePrefix = "arXiv",
    primaryClass = "hep-ph",
    reportNumber = "IFJPAN-IV-2018-07, BU-HEPP-18-03, MPP-2018-91",
    doi = "10.1016/j.physletb.2019.01.012",
    journal = "Phys. Lett. B",
    volume = "790",
    pages = "314--321",
    year = "2019"
}

@article{gribv-lptv:1972,
    author = "Gribov, V.N. and Lipatov, L.",
    title = "{e+ e- pair annihilation and deep inelastic e p scattering in perturbation theory}",
    journal = "Sov. J. Nucl. Phys.",
    volume = "15",
    number = "9",
    pages = "675, 938",
    year = "1972"
}

@article{Jadach:2023cl1,
      author         = "Jadach, S. and Ward, B. F. L. and  W\c{a}s, Z.",
      title          = "{Collinearly Enhanced Realizations of the Yennie-Frautschi-Suura (YFS) MC Approach to Precision Resummation Theory}",
      journal     = "Phys. Lett. B",
      volume     = "848",
      pages       = "138361",
      year           = "2024",
      eprint         = "2303.14260",
      archivePrefix  = "arXiv",
      reportNumber   = "BU-HEPP-23-01",
      SLACcitation   = "%%CITATION = ARXIV:2303.14260;%%",
}

@article{Jadach:2000ir,
      author         = "Jadach, S. and Ward, B. F. L. and  W\c{a}s, Z.",
      title          = "{Coherent exclusive exponentiation for precision Monte
                        Carlo calculations}",
      journal        = "Phys. Rev.",
      volume         = "D63",
      pages          = "113009",
      doi            = "10.1103/PhysRevD.63.113009",
      year           = "2001",
      eprint         = "hep-ph/0006359",
      archivePrefix  = "arXiv",
      primaryClass   = "hep-ph",
      reportNumber   = "CERN-TH-2000-087, UTHEP-99-09-01",
      SLACcitation   = "%%CITATION = HEP-PH/0006359;%%",
}

@article{Jadach:1999vf-sh,
      author         = "Jadach, S. and Ward, B. F. L. and  W\c{a}s, Z.",
      title          = "{The Precision Monte Carlo event generator K K for two
                        fermion final states in e+ e- collisions}",
      journal        = "Comput. Phys. Commun.",
      volume         = "130",
      pages          = "260-325",
      doi            = "10.1016/S0010-4655(00)00048-5",
      year           = "2000",
      reportNumber   = "DESY-99-106, CERN-TH-99-235, UTHEP-99-08-01",
      SLACcitation   = "%%CITATION = HEP-PH/9912214;%%",
}

@article{Jadach:2022vf,
      author         = "Jadach, S. and Ward, B. F. L. and  W\c{a}s, Z. and Yost, S.A and  Siodmok, A.",
      title          = "{Multi-photon Monte Carlo event generator KKMCee for lepton and quark  
                           pair production in lepton colliders}",
      journal        = "Comput. Phys. Commun.",
      volume         = "283",
      pages          = "108556",
      doi            = "10.1016/j.cpc.2022.108556",
      year           = "2023",
      eprint         = "2204.11949",
      archivePrefix  = "arXiv",
      primaryClass   = "hep-ph",
      reportNumber   = "IFJPAN-IV-2022-6, BU-HEPP-22-02, MCnet-22",
      SLACcitation   = "%%CITATION = ARXIV:2204.11949;%%",
}

@Article{ichep2022-say,
     author    = "{S.A. Yost} and others",
     title     = "{KKMChh; matching CEEX photonic ISR to a QED parton shower}",
     journal   = "PoS",
     volume    = "ICHEP2022",
     year      = "2022",
     pages     = "887",
     eprint     =  "2211.17177",
     archivePrefix  = "arXiv"
}

@Misc{latt3,
  author = 	 {Fodor, Z.},
  note = 	 {"$g-2$ from lattice QCD: the most recent result", talk in {\it ICHEP2024}},
  year =	 2024
}

@Misc{fcc2023wkshpms,
  author = 	 "Skrzypek, M. and others",
  note = 	 {talk in {\it 2023 FCC Workshop, Krakow, PL}},
  year =	 2023
}

@Article{ijmpa2018,
     author    = "{B.F.L. Ward}",
     title     = "{Prediction for the Cosmological Constant and Constraints on Susy GUTS in Resummed Quantum Gravity}",
     journal   = "Int. J. Mod. Phys. A",
     volume    = "33",
     year      = "2018",
     pages     = "1830028"
}

@Article{fabiola-1-10-23,
  author = "Gianotti, F. and others",
  note = "{in {\it Looking forward to the  New Year ...}, CERN, Geneva, Switzerland, Jan. 10, 2023}",
  year = 2023
}

@Article{grojean-2-13-24,
  author = "Grojean, C. and others",
  note = "{talk, CERN, Geneva, Switzerland, Feb. 13, 2024}",
  year = 2024
}

@misc{bflw-ichep2026,
  author = "Ward, B.F.L.",
  note = "{talk, in {\it ICHEP2026 Proceedings}, Aug. 1, 2026}",
  year = 2026
}

@Article{berends-neerver-burgers:1988,
  author = 	 {F.A. Berends and W.L. Van Neerven and G.J.H. Burgers},
  title = 	 { },
  journal = 	 {Nucl. Phys.},
  year = 	 {1988},
  volume =	 {B297},
  pages =	 {429; Erratum: Nucl. Phys. {\bf B304} (1988) 921 - 922}}

@Article{jad-yost-afb,
  author = {S. Jadach and S. Yost},
  title     = {QED Interference in Charge Asymmetry Near the Z Resonance at Future Electron-Positron Colliders},
  journal = {Phys. Rev. D},
  volume = {100},
  year     = {2019},
  pages   = {013002},
  eprint         = "1801.08611",
  archivePrefix  = "arXiv",
  primaryClass   = "hep-ph"
}

@Misc{sjetaltoappear,
  author = 	 {Jadach, S. and others},
  note = 	 {{to appear}},
  year =	" 2021"
}

@Misc{bflwetaltoappear,
  author = 	 {Ward, B. F. L. and others},
  note = 	 {{to appear}},
  year =	 "2023"
}

@Article{fjeger-fccwksp2019,
  author = "Jegerlehner, F.",
  journal = "{CERN Yellow Reports: Monographs, eds. A. Blondel {\it et al.}, CERN-2020-003}",
       year   = "2020",
       pages = "9",
       doi     = "10.23731/CYRM-2020-003",
  eprint       = "1905.05078",
      primaryClass   = "hep-ph",
       archivePrefix  = "arXiv"
}

@Article{gps:1998,
     title     = "Global positioning of spin GPS scheme for half spin 
                    massive spinors",
     journal   = "Eur. Phys. J.",
     volume    = "C22",
     year      = "2001",
     pages     = "423-430",
     eprint    = "hep-ph/9905452",
     archivePrefix = "arXiv",
     author    = "Jadach, S. and Ward, B.F.L. and W\c{a}s, Z."
}

@Article{atlas-epjc-zg:2024,
  author =	 {G. Aad and others},
  title =	 {Study of $Z\rightarrow \ell\ell\gamma$ decays at $\sqrt{s} = 8$ TeV in the ATLAS experiment at the LHC},
  journal = "Eur. Phys. J. C",
  volume = "84",
  year      = "2024",
  pages    = "195",
  eprint    = "arXiv:2310.11574",
  primaryClass = "hep-ex",
  SLACcitation  = "%%CITATION = arXiv:2310.11574;%%"
}

@Article{yfs:1961,
  author = 	 {Yennie, D. R. and Frautschi, S. and Suura, H.},
  title = 	 {{The infrared divergence phenomena and high-energy processes}},
  journal = 	 {Ann. Phys. (NY)},
  year = 	 "1961",
  volume =	 "13",
  pages =	 "379 - 452",
   doi     =   "10.1016/0003-4916(61)90151-8",
}

@Article{bhlumi4:1996,
     author    = "S. Jadach and W. Placzek and E. Richter- W\c{a}s and B. F. L.
                  Ward and Z.  W\c{a}s",
     title     = "{Upgrade of the Monte Carlo program BHLUMI for Bhabha
                  scattering at low angles to version 4.04}",
     journal   = "Comput. Phys. Commun.",
     volume    = "102",
     year      = "1997",
     pages     = "229",
     SLACcitation  = "%%CITATION = CPHCB,102,229;%%"
}

@Article{ceex2:1999sh,
  author = "Jadach, S. and Ward, B.~F.~L. and W\c{a}s, Z.",
  title = "{Coherent Exclusive Exponentiation For Precision Monte Carlo Calculations}",
  journal = "Phys. Rev. D",
  volume  = "63",
  year = "2001",
  pages   = "113009"
}

@Article{Jadach:2016kh,
     author    = "Jadach, S. and Ward, B. F. L. and  W\c{a}s, Z. and Yost, S.A.",
     title = {{${\cal KK}$MC-hh: Resummed Exact ${\cal O}(\alpha^2L)$ EW Corrections in a Hadronic MC Event Generator}},
     journal = "Phys. Rev. D",
     volume = "94",
     year = "2016",
     pages = "074006",
     eprint = "1608.01260",
     archivePrefix = "arXiv"
}

@Article{kkmchh1-sh,
     author    = "Jadach, S. and Ward, B. F. L. and  W\c{a}s, Z. and Yost, S.A.",
     title = {{Systematic Studies of Exact ${\cal O}(\alpha^2L)$ CEEX EW Corrections in a Hadronic MC for Precision Z/$\gamma^*$ Physics at LHC Energies}},
     journal = "Phys. Rev. D",
     volume = "99",
     year = "2019",
     pages = "076016"
}

@Article{kkmchh2,
     author    = "Jadach, S. and Ward, B. F. L. and  W\c{a}s, Z. and Yost, S.A.",
     title = {{IFI and ISR Effects for $Z/\gamma*$ Drell-Yan Observables using {\cal KK}MC-hh}},
     year = "2020",
     eprint = "2002.11692",
     archivePrefix = "arXiv"
}

@Article{mcnlo-hwiri,
     author    = "{ S.K. Majhi} and others",
     journal = "  Phys. Lett. B",
     volume = "719",
     year = "2013",
     pages = "367",
     eprint = "1208.4750",
     archivePrefix = "arXiv"
}

@Article{mcnlo-hwiri1,
     author    = "{ A. Mukhopadhyay and B.F.L. Ward}",
     journal = " Mod. Phys. Lett. A",
     volume = "31",
     year = "2016",
     pages = "1650063"
}

@Article{mcnlo,
     author    = "{S. Frixione and B.Webber}",
     journal = " J. High Energy Phys.",
     volume = " 0206",
     year = "2002",
     pages = "029"
}

@Article{mcnlo1,
     author    = "{ S. Frixione} and others",
     journal = " J. High Energy Phys.",
     volume = " 1101",
     year = "2011",
     pages = "053",
     eprint = "1010.0568",
     archivePrefix = "arXiv"
}

@Article{frixione-2021,
     author    = "{F. Frixione}",
     journal = " J. High Energy Phys.",
     volume = "2021",
     year = "2021",
     pages = "180",
     eprint = "2105.06688",
     archivePrefix = "arXiv"
}

@Article{frixione-2019,
     author    = "{F. Frixione} ",
     journal = " J. High Energy Phys.",
     volume = "1911",
     year = "2019",
     pages = "158",
     eprint = "1909.03886",
     archivePrefix = "arXiv"
}

@Article{bertone-2019,
     author    = "{V. Bertone} and others",
    journal = " J. High Energy Phys.",
     volume = "2003", 
     year = "2020",
     pages = "135",
     eprint = "1911.12040",
     archivePrefix = "arXiv"
}

@misc{bertone-2022,
     author    = "{V. Bertone} and others",
     title = "{Improving methods and predictions at high-energy $e^+e^−$ colliders within collinear factorisation}",
     year = "2022",
     eprint = "2207.03265",
     archivePrefix = "arXiv",
     note = "TIF-UNIMI-2022-10, ZU-TH 26/22"
}

@Article{latt2,
     author    = " Ce, M. and others",
     title = {The hadronic running of the electromagnetic coupling and the electroweak mixing angle from lattice QCD},
     journal = "J. High Energy Phys.",
     volume = "08",
     year = "2022",
     pages = "220",
     eprint = "2203.08676",
     archivePrefix = "arXiv"
}

@Article{latt1,
     author    = "Borsanyi, Sz. and others",
     title = {Hadronic vacuum polarization contribution to the anomalous magnetic moments of leptons from first principles},
     journal = "Phys. Rev. Lett",
     volume = "121",
     year = "2018",
     pages = "022002",
     eprint = "1711.04980",
     archivePrefix = "arXiv"
}
\bibliographystyle{utphys_spires}

\end{document}